\documentclass[12pt]{article}

 \usepackage{amsfonts}
 \usepackage{amssymb,amsmath}
 \usepackage{graphicx} \usepackage{epstopdf}
 \newcommand{\crlb}[1]{\label{#1}\\[2pt]}
 
 \newcommand{\crld}[1]{\label{#1}}
 \newcommand{\eela}[1]{\quad\hbox{\scriptsize{#1}}\label{#1}\end{eqnarray}}
 \newcommand{\eelb}[1]{\label{#1}\end{eqnarray}}
 
 \newcommand{\newsecb}[2]{\section{#1}\label{#2}\setcounter{equation}{0}}

 \newcommand{\nolabels} {\def\eel{\eelb}\def\eeql{\eeqlb}  \def\crl{\crlb} \def\newsecl{\newsecb}\def\bibiteml{\bibitem} \def\citel{\cite}\def\labell{\crld}}
\newcommand{\eeqla}[1]{\quad\hbox{\scriptsize{#1}}\label{#1}\end{aligned}\end{equation}}
\newcommand{\eeqlb}[1]{\label{#1}\end{aligned}\end{equation}}

\newcommand\publishversion{\nolabels\setlength{\textheight}{8.3in}\setlength{\oddsidemargin}{0in}
   	 \setlength{\textwidth}{6.3in}\setlength{\topmargin}{-0.2in}}

\def\beq{\begin{equation}\begin{aligned}}		\def\eeq{\end{aligned}\end{equation}}
\def\be{\begin{eqnarray}}  					\def\ee{\end{eqnarray}}		
\def\bi#1{\begin{itemize}\item[#1]} 			 			\def\ei{\end{itemize}} 
  \def\eqn#1{(\ref{#1})}
   	 \def\fn{\footnote}	  		\def\nm{\nonumber} 

		    \def\b{\beta}   \def\d{\delta}   \def\k{\kappa}     \def\l{\lambda}  
		       \def\r{\varrho}      \def\t{\tau}    
    		\def\G{\Gamma}  	\def\D{\Delta}    
\def\m{\mu}	    		        		     		\def\vv{\varphi}     \def\n{\nu}
 	 		\def\s{\sigma}     	\def\tht{\theta}      	 
	     		          		              	  
\def\W{\Omega}    		  		\def\dd{{\rm d}} 		  
\def\OO{{\mathcal O}} 		     
 		      
\def\pa{\partial}			\def\ra{\rightarrow}	
\def\bra{\langle} 		\def\ket{\rangle}

	\def\iss{\ =\ }
 
\def\fract#1#2{{\textstyle\frac{#1}{#2}}}	 	 \def\fractje#1#2{{\scriptstyle\frac{#1}{#2}}}	
\def\ffract#1#2{\raise .2 em\hbox{$\scriptstyle#1$}\kern-.3em/\kern-.2em\lower .15 em \hbox{$\scriptstyle#2$}}
\def\half{\fract12}					\def\halfje{\fractje12}
\def\tl#1{\tilde{#1}} 
\def\ex#1{e^{\textstyle#1}} 		\def\qqquad{\qquad\qquad}		
\def\bpmatrix{\begin{pmatrix}} 			\def\epmatrix{\end{pmatrix}}
\def\bmatrix{\begin{matrix}} 			\def\ematrix{\end{matrix}} 
\def\bcenter{\begin{center}}			\def\ecenter{\end{center}}

\def\lowerheightfig#1#2#3{\(\raise-#1\hbox{\includegraphics[height=#2]{#3}}\)}
\def\lowerwidthfig#1#2#3{\(\raise-#1\hbox{\includegraphics[width=#2]{#3}}\)}
		 
\def\weglaten#1{}		\def\ul#1{\underline{#1\vphantom{p}}}
\def\BH{{\mathrm{BH}}}\def\Pl{{\mathrm{Pl}}}\def\inn{{\mathrm{in}}} \def\outt{{\mathrm{out}}}
\def\ds{\displaystyle}
\publishversion

\begin{document}

\begin{titlepage}
 \title{{ \LARGE\textbf {The Quantum Black Hole as\\[5pt] a Theoretical Lab\\[10pt]}}
	{\large\textbf{ A pedagogical treatment of a new approach}}}
		\author{Gerard 't~Hooft}
\date{\small  Institute for Theoretical Physics \\ Utrecht University  \\[10pt]
 Postbox 80.089 \\ 3508 TB Utrecht, the Netherlands  \\[10pt]
e-mail:  g.thooft@uu.nl \\ internet: 
http://www.staff.science.uu.nl/\~{}hooft101/  \\[10pt]
{Lectures held at the International School of
	Subnuclear Physics,\\ \textsc{``FROM GRAVITATIONAL WAVES to QED, QFD and QCD"},\\
	``Ettore Majorana" Foundation and Centre for Scientific Culture,\\June 2018}\\[10pt]
}
 \maketitle
\noindent \textbf{Abstract}\\ [10pt]
	Getting the mathematical rules for quantised black holes correctly is far from straightforward. Many earlier treatises got it not quite correctly. The general relativistic transformation linking the distant observer (who only detects particles outside the hole) with the observations of a local observer (who falls into a black hole) must map the quantum states in a one-to-one way. This does not come out right if one follows text book rules. Here it is advocated that demanding very strict logic leads to new insights, such as the non-triviality of space-time topology near a black hole. This way one may attempt to make up for the lack of direct experimental evidence concerning gravitation at the Planck scale.
	It is noted that this approach does not require assumptions such as string theories or AdS/CFT conjectures. All we need to assume is the validity of quantum field theory wherever the Schwarzschild metric is regular, combined with the requirement that only those general coordinate transformations apply that map pure quantum states one-to-one onto pure quantum states.

 \end{titlepage}							
  \setcounter{page}{2}
  \newsecl{Introduction}{intro}
  
While mysteries still abound,  many discoveries have been made concerning
the world of the most fundamental processes between elementary particles of matter, their laws of motion and their ways of interacting, together with the laws controlling the nature of space and time themselves, while the universe came into being and evolved further. At one side of the scenery we have experimental scientists and observers, while at the other side the theoreticians attempt to see order and systematics in what has been found, allowing them to point at conceivable new phenomena and regularities that the observers and experimentalist could try to investigate.
  
On the one hand, it seems that, today, a new stage of understanding has been reached that we can all be proud of: the \emph{Standard Model} appears to work very well\,\cite{SM}. Yet there are still numerous mysteries that, by their nature, resemble the mysteries of the 1960s, while others seem to be much more hopeless. While we are not giving up on sharpening our understanding and insights, we also realise that making real progress is becoming gradually more difficult. 

A good rule used to be that we simply have to look at those spots in our theories where existing explanations are wanting, where competing approaches contradict one another, and where predictions are difficult to compute. New theories are constructed, bits and pieces are combined, and, after elaborate experimentations and observations, the right answers are selected.

But, in the science of the most fundamental interactions of matter, this simple description of our activities does not quite apply anymore. Where new theories could really make a difference, at energies billions of times higher than what is presently available, and when gravitational interactions between particles come into play, we cannot really do experiments anymore. What is done instead is that arbitrary, usually extremely naive, theoretical ideas are brought forward, coming in large numbers, while the competition depends on the fad of the day, and decent selection mechanisms are almost absent. We cannot blame the scientists themselves for this; it is simply a fact that Nature does not provide us with more direct clues, and the situation we are in now is actually a result of this fact: theories that could be checked with the means we have today, have gone through appropriate selection mechanisms and are now either pars of the text books or they are forgotten. What is left to be investigated are the theories and principles that we still have not been able to differentiate between.

In this lecture, we advocate that we can de better, but the differentiation between good and bad theories will require a very precise and thorough analysis. The quality of different theoretical procedures can be compared and improved; we advocate to rely on one piece of experience that was quite compelling in the past: Nature works along lines demarcated by extremely strict logical rules. The best theories are the ones that leave no doubts as to what they say, what the rules are for their calculations, and what the conditions are for their applicability.

As stated above, the gravitational force must eventually play a central role in the fundamental theories. All particles and other forms of matter interact gravitationally according to very strict logical rules. Whenever the test particles to which we apply our theories of gravity, are sufficiently accurately described by classical laws of mechanics, we see that Einstein's theory of General Relativity applies, and what these laws say is utterly clear. Under very precisely defined circumstances, black holes must form. Black holes have been studied extensively, for most parts purely theoretically, while very recently also independent experimental observations can be added to what we know: the LIGO and VIRGO observations of gravitational waves indicate that the classical theoretical insights represented by General Relativity work.\fn{See the lectures by Profs. R. De Salvo and W. Del Pozzo  at this school.}  Large black holes behave exactly as expected.
  										
However, at the tiniest distance scales, particles do not behave classically but quantum mechanically. This will require that we should also insert the laws of quantum mechanics in our descriptions of black holes. Black holes always appeared to behave like pieces of matter, characterised by mass, angular momentum and charge(s), and their laws of behaviour are in agreement with classical (relativistic) mechanics, but now, what has to be done, is replace these laws by the appropriate quantum mechanical ones. It is here that we emphasise our appeal to the application of stricter logic than usual.							

Candidate theories do exist. Most notably, (super)string theory\,\cite{sustr} is often resorted to. But there is a logical gap to be worried about. String theory was supposed to take over when particle energies and momenta approach the Planckian domain.  But the particles and fields that seem to be involved with black holes with mass \(M_\BH\), all seem to have energy excitations in the domain \(M_\Pl^2/M_\BH\) in natural units, where \(M_\Pl\) is the Planck mass. These energies are way below the Planck domain. It should be possible to describe the properties of black holes with mass \(M_\BH\gg M_\Pl\) without the use of string theory at all. We now claim that the assumptions ``\emph{particles are pieces of string}", and black holes are ``\emph{stacks of \(D\)-branes}"\,\cite{Maldacena}, that can be subject to the AdS/CFT conjectures, are better to be avoided in the case of  large, heavy black holes. These theories have not yet passed the selection mechanism of experimental observation.

How far can we get in describing the physics of black holes, purely by applying effective perturbative gravity, combined with Standard Model physics, all at energies \(\ll M_\Pl\), using only standard quantum mechanics and General Relativity?\fn{These theories are also assumptions, but of course they are quite compelling. Dropping or loosening these assumptions would be much more difficult  than keeping them as long as we can. No theory is sacred, and some sense of taste for what to keep and what not, is a useful attire for a theoretical physicist.}

We claim that topics such as \emph{microstates}\,\cite{Bek1, Bek2} can be understood very well in such an approach. Counting the microstates is still difficult\,\cite{GtHmicro}, as not all features of this theory have been extensively explored. There are several extremely powerful tools that must be used here. A consequence of the fact that some of our equations where the gravitational force is included are nevertheless linear, can be exploited by performing an expansion of the physical data in \emph{spherical harmonics}. The different spherical modes then decouple. The equations factorise, a situation very reminiscent of the Schr\"odinger equation for the hydrogen atom\ \cite{GtHdiagonal}. Its quantum states can be labelled in terms of the quantum numbers \(\ell\) and \(m\). We here have a similar situation\fn{Note that, in our work, \(\ell\) and \(m\) do \emph{not} refer to individual quantum states, but to spherical modes of position and momentum operators.}. As in the hydrogen atom, the equations of matter and its gravitational interactions reduce to 1+1 dimensional partial differential equations, which are very simple to solve.

One then gets an embarrassingly simple picture of what is failing in the older theories of black holes: \emph{in those theories, quantum states are not counted the way they should be; quantum ``cloning" should be inhibited.} In older theories, this problem was signalled, but not appropriately understood. What has to be done is now standing out, as will be explained: we have always been working in the wrong space-time manifold.

Thus we argue that the quantum black hole can be regarded as a ``theoretical laboratory". It replaces the experimental tests that would have been urgently needed to understand quantum gravity. We use stricter logic instead.\fn{Or, we are trying to. Heated discussions are going on as to what the correct attitude should be; no consensus has been reached yet.}
							
At first sight, it will seem that what we do is plainly wrong, and it was criticised as such in email exchanges with colleagues. But the equations are so simple that one can now investigate all alternatives. It is essential that we use Cauchy surfaces, and examine how the data on these surfaces evolve with time. What is the time coordinate? How can we ensure that the Cauchy data evolve through unitary evolution operators?

The danger is that the Cauchy data get lost between the crevices of the horizons, which can easily happen while you think you are doing things correctly. \emph{What would a local observer, falling through the horizon, see?} Our experience is that the expansion in spherical harmonics does here what experiments did when physicists were constructing the Standard Model. Thus we advocate the use of spherical harmonics, or whatever else we can put our hands on, wherever we can, so as to obtain a clearer picture of our subject.

\newsecl{The tortoise coordinates}{tortoise}
Our prototype is the Schwarzschild black hole. No serious complications are expected when these investigations are generalised to Kerr-Newman\,\cite{Newman} black holes. In string theories one often considers the \emph{extreme limit} of Kerr-Newman or Reissner Nordstr\"om black holes, but in these limiting cases the horizon does turn into something different from what it was in the generic case; since there, the temperature tends to zero, these limiting cases are often difficult to reach physically; we did not study what might happen there.

Writing \(M_\BH=M\), the Schwarzschild metric is:
	\beq &&\dd s^2 \ =\  g_{\m\n}\dd x^\m\dd x^\n\ ,\qquad\hbox{where \(g_{\m\n}\) can be read off from}
	\qquad\qquad&\nm\\ 
	\dd s^2 &=& \frac 1{1-\fract{2GM}r}\,\dd r^2-\big(1-\fract{2GM}r\big)\,\dd t^2+r^2\dd\W^2\ ;\quad 
	\left\{\begin{matrix}\W&\equiv&(\tht,\,\vv)\ ,\qquad\\[3pt] \dd\W&\equiv&(\dd\tht,\ \sin\tht\dd\vv)\ .\end{matrix}
	\right. & \eeql{Schwsch}
The Kruskal-Szekeres coordinates \(x,\ y\) are defined by
	\beq  x\,y&\ \ =\ \ \left(\frac r{2GM}-1\right)\,e^{\,r/2GM} ; \\[3pt]
	 x/y&\ \ =\ \  e^{\,t/2GM}\quad. \\[3pt]
	 \dd s^2&\ \ =\ \ \frac{32(GM)^3}r\,e^{-r/2GM}\,\dd x\,\dd y+r^2\dd\W^2\ .  \eeql{Kruskal}
At \(r=2GM\), we have the past event horizon at \(x=0\) , and the future event horizon at  \(y=0\) .

Let us now concentrate on the region close to the horizon, \(r \approx 2GM\) . There we write
\be x=\frac{\sqrt{e/2}}{2GM}\,u^+\ ,\qquad y=\frac{\sqrt{e/2}}{2GM}\,u^-\ ;\qquad 2GM\equiv R\ ,\ee 
 so that space-time looks approximately flat, with \(u^\pm\) acting as light cone coordinates:
\be\dd s^2\ra 2\dd u^+\dd u^-+R^2\dd\W^2\ . \ee
Rescaling the Schwarzschild time, \(t/4GM=\t\ ,\) we find that, in a geodesic, the coordinates \(u^\pm\)  scale with time as follows:
\be u^-(\t) = u^-(0)\,e^{\,\t}\ ,\quad\hbox{and}\quad u^+(\t) = u^+(0)\,e^{-\t}\ . \eel{timescale}
The relation between the coordinates is sketched in Fig.~\ref{Szekeres}. As time \(\t\) advances, its \(u^+\) coordinate approaches the future event horizon, while the \(u^-\) coordinate goes to the past event horizon (see arrow). The straight tilted lines are the equal time lines \(\t\) for the distant observer using the Schwarzschild coordinates. The in-going particles generate wave functions defined on the \(u^+\) axis, particles going out originate as wave functions on the coordinate \(u^-\).

\begin{figure} \begin{center}
 \lowerwidthfig{150pt}{180pt}{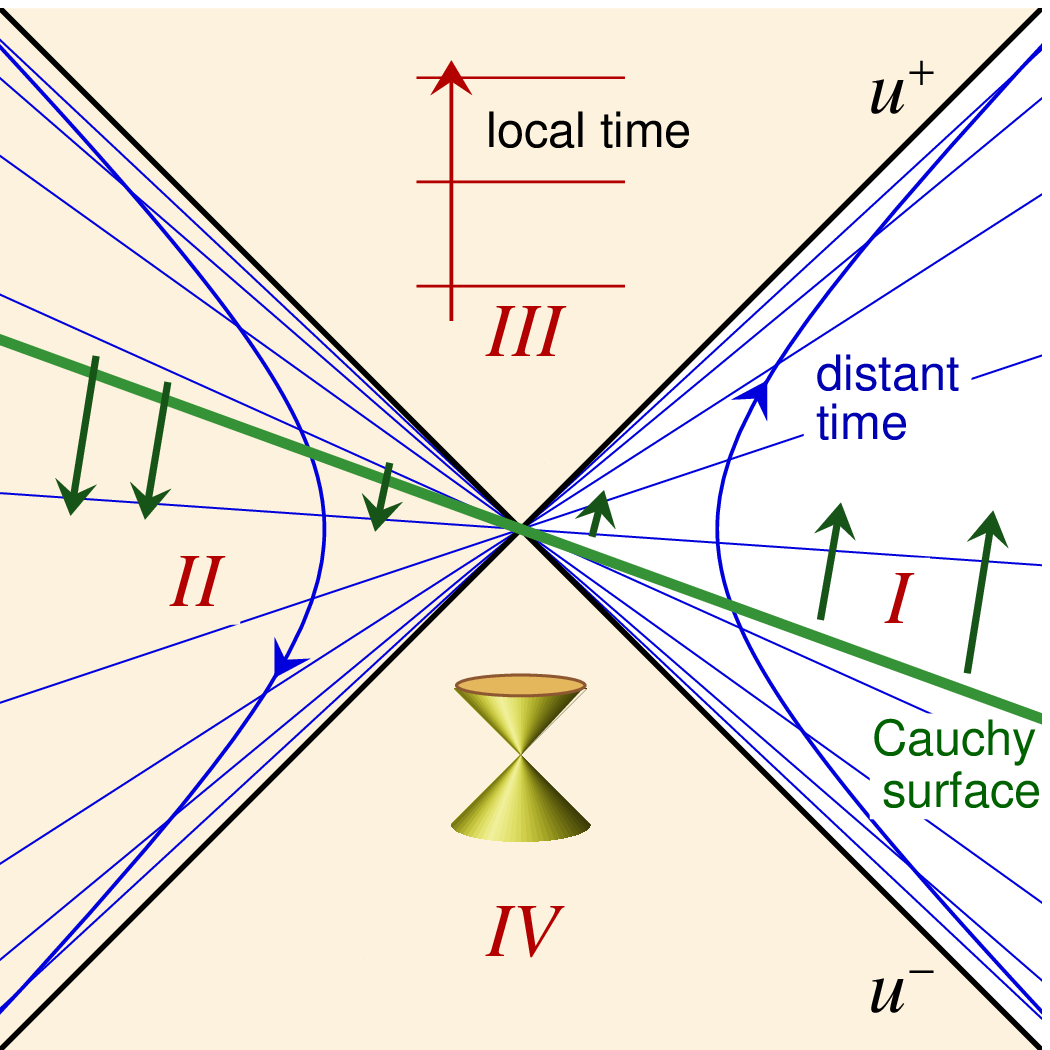} \quad \begin{caption}
 {\small The Kruskal-Szekeres coordinates near the future and past event horizons. Conventionally, region \(I\) is considered to be the visible part of the universe, the shaded parts, \(II,\  III,\) and \(IV\) are interior regions, from which no information can escape. Everywhere, the light cone is oriented as shown.} \labell{Szekeres}
 \end{caption} \end{center}
\end{figure}

A useful way to display the space-time topology of a black hole, is the Penrose diagram, obtained by the replacements
\be u^+\ra f^+(u^+) \quad \hbox{ and } \quad u^-\ra f^-(u^-)\ , \eel{penroseupum}
such that the new coordinates \( f^\pm \) are both contained in the segment \( [-1,1] \).
The metric keeps the form 
\be \dd s^2=2A(u)\dd u^+\dd u^-+r^2(u)\,\dd\W^2\ , \eel{penrosemetric}
see Fig.~\ref{Penrose}. Depicted is the case of the \emph{eternal black hole}, which means that we neither include the effects of the imploding matter that forms the black hole, nor those of the evaporated matter that might end its existence at much later times. We have good reasons for this: Since the implosion and the final evaporation take place in the very far past or the very far future, they do not affect what we see in a more modest time period \([-T,\,+T]\), as long as \(T\ll\OO(M^3_\BH)\) in Planck units.
Do note that, in the diagram, the regions \(III\) and \(IV\) are, in a sense, unphysical, because the outside, distant observer has no time coordinates for them; the external time for these regions is way beyond the time stretch that seems to cover eternity, \([-\infty,\,+\infty]\).

The reason for including region \(III\) is that an observer falling in would not be able to exactly locate the position of the future event horizon, so for this observer, the laws of physics apply in an extended region of space-time. Since this observer cannot detect the Hawking particles, it is legitimate to omit these entirely, so from that point of view, region \(III\) is physical.

But the laws of physics considered so-far, with or without quantum mechanics, are time-reversible. So if it is legal to keep region \(III\), it may also be legal to consider slight extrapolations to region \(IV\): \emph{an observer going out together with the Hawking particles, cannot observe the imploding matter.}

\begin{figure} \begin{center}

 \lowerwidthfig{150pt}{250pt}{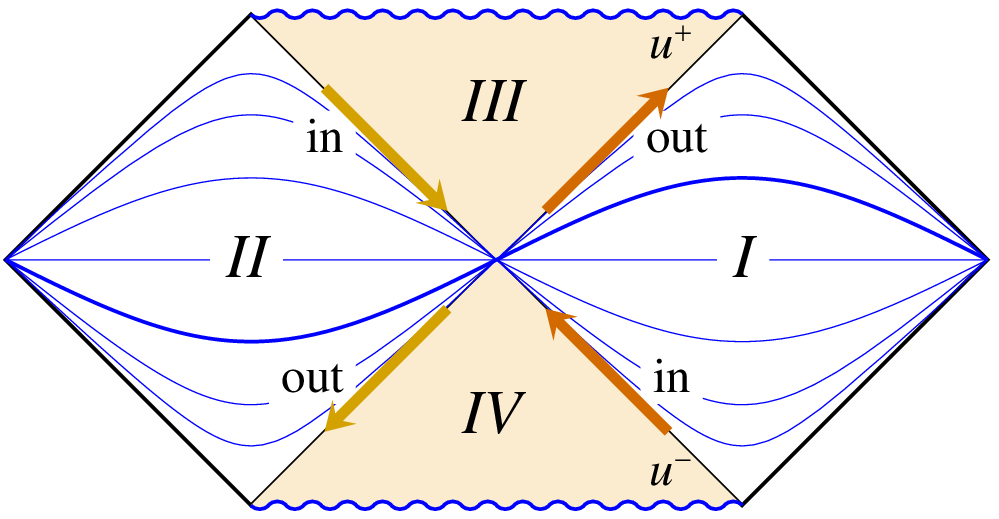} \quad \begin{caption}
 {\small The Penrose diagram, obtained by mapping \(u^+\) onto a variable 
 on the segment \([-1,1]\) and similarly \(u^-\). Equal-time lines (for a distant observer) are shown. The solid curve is a possible Cauchy surface.} \labell{Penrose}
 \end{caption} \end{center}
\end{figure}

\newsecl{The microstates}{microstates}

The procedure sketched above puts us in the position to describe the black hole microstates. These are the states a black hole will be in when surrounded by in- and out-going elementary particles. These particles are only considered during a short time interval \([-T,\,T]\), preferably with \(T\approx M_\BH\)  in Planck units. We emphatically exclude larger objects falling in. Assuming the general idea of Hawking radiation to be correct, at least qualitatively,  the probability for large objects emerging from the black hole will be immensely suppressed by their thermal Boltzmann factor, \(e^{-\b_H E}\) in Planck units (where \(\b_H\) is the inverse Hawking temperature, which is of order \(M_\BH\), and \(E\gg M_\Pl\) is the mass of the heavy in-falling object.).

All particles going in and out, are considered at a given moment in time, which means they are represented by (quantum) states defined on a Cauchy surface. We \emph{exclude} all those particles that arrived much earlier, or leave at a much later time. These particles will be squeezed against one of  the horizons at exponentially small distance, while their kinetic energies, as experienced by an observer on this Cauchy surface, have grown to exponentially large values. As was stated at the end of the previous section, these particles will as yet not be included in our description, because they would cause new sources of curvature on space-time that would invalidate the entire description of the Penrose diagram.

Now this can only be a temporary, intermediate, stage of our discussion. We use it as a starting point to describe small variations in the black hole state during small periods in time. Later, we will address the black hole{'}s long-term time dependence more accurately. The problem is of course: as time goes by, particles that were innocent a few moments ago, will squeeze against the horizon. How do we take care of them, if these very energetic particles are supposed to be excluded? In short, if we leave these particles out, one will encounter evolution operators that violate unitarity. How to cure this apparent unitarity violation will now be discussed. 

The notion of ``hard" and ``soft" particles was used by other authors in a way that differs from ours. A ``soft" particle is here taken to be a particle whose mass \(\m\) and momentum \(\vec p\) are all taken to be small compared to the Planck mass, \(M_\Pl\). On the Cauchy surface, our particles will be on the mass shell, 
\(2p^+p^-+\tl p^2+\m^2=0\). Here \(|\tl p|\approx L/R\ ,\ \m=\) mass, and the longitudinal momenta depend on time as
\be p^-(\t)=p^-(0)e^\t\ ,\quad p^+(\t)=p^+(0)e^{-\t}\ . \quad \tl p\ \hbox{ and }\ \m\ \hbox{are constant in time.} 
\eel{timedependence}
A particle is \emph{soft} if \(|\vec p\,|,\ \m\ll M_\Pl\ ,\) in which case its effects on space and time are negligible, 
and it is \emph{hard} otherwise. All particles that are soft at a given time \(t\) will have \(p^+\ra 0,\ p^-\ra\infty\) at much later times, and had \(p^-\ra 0,\ p^+\ra\infty\) at much earlier time. Clearly, to understand what happens with the evolution at longer time intervals, we have to understand what particles do when they turn into hard particles.

At \(\t\gg 1\), all in-particles become hard, as \(p^- \ra\infty\). Their interactions with other in-particles are negligible (they basically move in parallel orbits), but they do interact with the out-particles. The interaction through QFT forces stay weak, but the \emph{gravitational forces} make that (early) in-particles interact strongly with (late) out-particles. This force may not be ignored. And, most importantly, its effects can be calculated.

Consider a hard in-particle, having a large value for \(p^-\). In its own reference frame, its gravitational field is very weak, and well described by the Schwarzschild metric, with a very tiny value of the mass \(\m\). One can now easily calculate what this gravitational field transforms into, in an other reference frame. In the reference frame where \(p^-\) gets large, one finds that space-time is still flat at points where \(x^+\) is positive, and at points where \(x^+\) is negative, but we have curvature where \(x^+\approx 0\). There, the two flat half-spaces are glued together with a mismatch \(\d x^-\) that is calculated to be\,\cite{aichelbsexl,GtHDray}
\be \d x^-(\tl x)=-4G\,p^-(\tl x)\,\log|\tl x-\tl x'|\ ,\eel{flatshift}
where \(\tl x\) are the transverse coordinates of the out-going particle, and \(\tl x'\) the transverse coordinates of the hard in-particle.

This curvature is taken into account elegantly if we replace this space-time curvature by a new law of physics: the soft \emph{out}-particles, going in the \(x^+\) direction, are \emph{dragged along} by an amount \(\d x^-\).
This is the effect registered by a local observer. It may still be tiny, while a distant observer will see some drastic effects, as sketched in Fig.~\ref{BHshift}. Because of this increase of the effect being exponential in time, this is often interpreted as a sign of `chaos' taking place\,\cite{ER=EPR}. However, we shall see that it is not as chaotic as one might fear; explicit calculations are still possible, and at the end the outcome is quite orderly. the evolution operator can still be computed, and, over time scales not exceeding the Schwarzschild time scale of the order of \(M_\BH\) in natural units, this matrix even factorises.

This gravitational effect of in-going particles on the out-going ones will be referred to as their gravitational \emph{footprint}. Eventually, we shall interpret all quantum data of the out-going particles as being footprints of the in-going ones, so that information going in is indeed preserved in the out-going particles.

\begin{figure} \qqquad\qqquad\qqquad
\lowerwidthfig{150pt}{150pt}{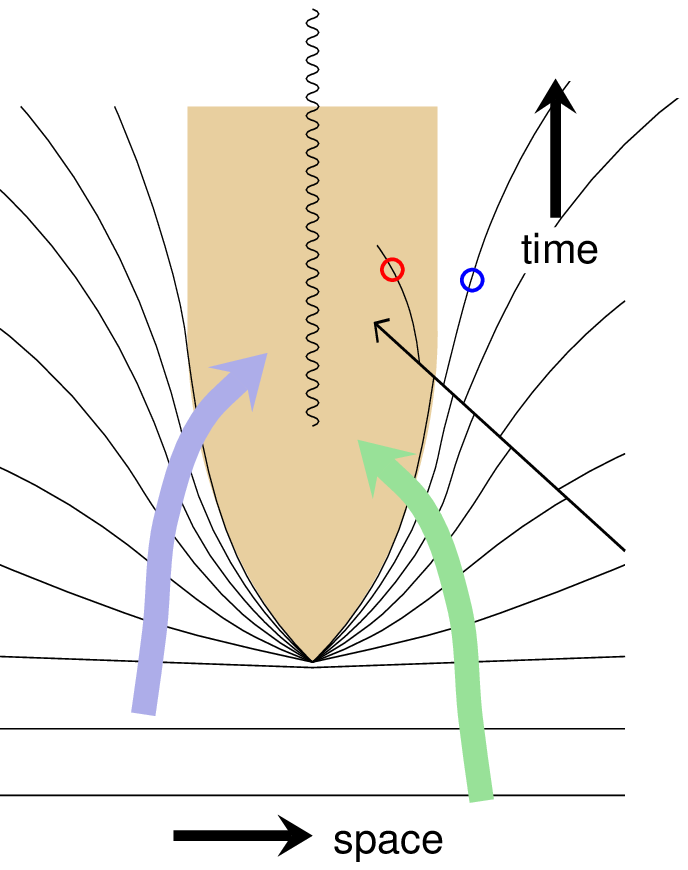} \quad \begin{caption}
{ \small The gravitational drag effect of an in-going particle (black arrow), an artist impression, still showing the effects of the initial implosion (broad arrows) and the onset of the future event horizon at an early time. One of the Hawking particles, on its way out, is dragged back in.}
 \labell{BHshift}
 \end{caption} \end{figure}

A remarkable fact is that, in our subsequent calculations, all that is needed is the interactions between hard particles and soft ones, or soft ones mutually (as described in the Standard Model Lagrangian in regions \(I\) and \(II\)).  Direct interactions between hard particles and other hard particles will \emph{not} have to be considered.

Thus, we start with only soft particles in the Cauchy data of a black hole --- the hard particles will be replaced by the footprints they leave behind in the spectrum of soft particles, as we shall see. It is now our task to calculate the evolution matrix, and investigate its unitarity in the Hilbert space spanned by soft particles only.

\newsecl{The  expansion in spherical harmonics}{sphericalharm}
	The gravitational dragging effect that hard particles have on soft particles, sometimes referred to as the Shapiro delay, was calculated first in a flat background\,\cite{aichelbsexl}, but the calculation can easily be extended to the background of a black hole horizon\,\cite{GtHDray}. In the Kruskal Szekeres coordinates \eqn{Kruskal}, the Shapiro shift \(\d u^-\) at solid angle \(\W=(\tht,\,\vv)\), caused by an in-going particle with momentum \(p^-\) at solid angle \(\W'=(\tht',\ \vv')\), is given by
	\be \d u^-(\W)=8\pi G\,f(\W,\,\W')p^-\ ;\quad (1-\D_\W)f(\W,\,\W')=\d^2(\W,\,\W')\ .\eel{shifthor}
If we have many in-going particles, the shift they cause is the sum of the individual shifts:
\be p^-(\W)=\sum_ip_i^-\d^2(\W,\,\W_i)\ ;\quad \d u^-(\W)=8\pi G\int\dd^2\W'\,f(\W,\,\W')\,p^-(\W')\ . 
	\eel{combshift}

A distant observer will see an unending stream of particles going in and out. The local observer may see a state close to his vacuum, or with soft particles added to that. One may assume a distribution \(p^-(\W)\) that represents all particles that went in during the entire history of the black hole.  This may even include the imploding matter that formed the black hole in the distant past.

The out-going particles will then be characterised by their positions \(u^-(\W)\) when they leave the past event horizon.\fn{Some authors assume the Hawking particles to emerge from the future event horizon by fluctuating faster than light. We find such a view difficult to maintain. An observer always sees particles going in through the future event horizon and particles going out through the past horizon.}
 Thus we write
 \be u^-_\outt(\W)=8\pi G\int\dd^2\W'\,f(\W,\,\W')\,p^-_\inn(\W')\ . \eel{u-inp-out}
Note that \emph{very early} in-going particles carry exponentially large values for \(p^-\). By including their contributions we encounter particles going out with hugely varying values of \(u^-_\outt\). This means that many of the out-going particles will emerge at much later or much earlier times \(\t\). When looking at a black hole at given time \(\t\), one only cares about those particles that are visible at finite positions in the relevant section of the Penrose diagram, and we ask how their positions are affected by in-going particles. In more complete treatments of this material we see how all these features can be represented in the appropriate equations.
\def\spc{\hbox{\qquad}}
For now, let us expand both \(u^\pm(\W)\) and \(p^\pm(\W)\)  in spherical harmonics, to find:	
\be \ds u^\pm(\W) =  \sum_{\ell,m}u_{\ell m}Y_{\ell m}(\W)\ , \  &&\spc  p^\pm(\W) =\ \sum_{\ell,m}
	p^\pm_{\ell m}Y_{\ell m}(\W)\ ; \labell{ellmdef} \\
 \ds [u^\pm(\W),\,p^\mp(\W')] = i\d^2(\W,\,\W')\ ,\ \spc &&    \ [u^\pm_{\ell m},\,p^\mp_{\ell'\,m'}]
	= i\d_{\ell\ell'}\,\d_{mm'}\ ;\labell{ellmcomm}	\\
 \ds  u^-_{\ell m,\,\outt} = \frac{8\pi G}{\ell^2+\ell+1}\,p^-_{\ell m,\,\inn}\ , &&\ \spc   u^+_{\ell m,\,\inn}
	= -\frac{8\pi G}{\ell^2+\ell+1} \, p^+_{\ell m,\,\outt}\ .	\eel{ellmshifts}
	
\noindent Here, \(p^\pm_{\ell m}\ =\ \) the total momentum of the  \(^{\outt}_{\inn}\)-particles in the (\(\ell,m\))-mode,
	\\ and  \  \(u_{\ell m}^\pm\ =\ \) the (\(\ell,m\))\,-\,component of the c.m. position of the 
	\ \(^{\inn}_{\outt}\)-particles.
	
Now comes a very essential observation: in the equations \eqn{ellmdef}---\eqn{ellmshifts}, the different \((\ell,\,m)\) modes  decouple completely.\fn{Some cross talk between different \((\ell,\,m)\) modes is expected at the highest values of \(\ell\) (when \(\ell=\OO(M_\BH)\) in Planck units), due to the transverse components of the gravitational interactions; these would otherwise be negligible.} So, in every \((\ell,\,m)\) mode there is exactly one variable \(p^-=p^-_{\ell,m,\,\inn}\) and one \(p^+\), while \(u^\pm\) are related to these by the last line of Eq.~\eqn{ellmshifts}. This means that we can readily solve the equations. What was thought to be a chaotic system now turns out to be integrable: the operators \(u^+\) for the in-particles are are just the Fourier complements of the operators \(u^-\) for the out-particles: \(p^-=-i\pa/\pa u^+\).
	
There are several further essential steps to be taken. The most difficult one is to sort out the relation between the momentum variables \(p^-_{\ell m}\) in the spherical harmonic modes and the in- and out-going particles as seen by the distant observers. In principle, all we have to do is to calculate the momentum distributions of all in-going particles. This gives the  positions \(u^-_{\ell m}\) of the out-going particles, which give us the momenta by Fourier transformations. But how do we recover the Fock space states of the soft out-going Standard Model particles from that? This question is both conceptually and technically difficult. For one thing, the vacuum state for the distant observer is not the vacuum state for the local observer, due to the required Bogolyubov transformation to go from one to the other. Some of the ensuing issues have not yet been explored completely.

Secondly, it is essential to note that the Penrose diagram shows \emph{two} unrelated asymptotic domains. It seems as if the black hole metric represents two universes \(I\) and \(II\), both containing the same black hole. This is sometimes thought to mean that we have two ``entangled" black holes, each sitting in some different place in either the same or different universes\,\cite{ER=EPR}. This author was unable to fathom what this means physically. If indeed the Penrose diagram stands for two black holes we have a deep and important problem: these two black holes interact. If it is true that the out-going particles are the Fourier transforms of the in-going ones, then annihilating one in-going particle in region \(I\), would entail the annihilation of an out-going particle that entangles regions \(I\) and \(II\). The creation of an in-going particle in one black hole would cease to commute with the annihilation of an out-going particle in region \(II\). In other words, these two black holes interact with one another, something that would be forbidden by any locality condition. A signal would go directly from one black hole to the other. That cannot be right.

A better solution of this problem is the assumption that regions \(I\) and \(II\) refer to the same black hole. However, they cannot refer to the same spot on the horizon. It so happens that there exists exactly one mathematical constraint that does the job correctly: \emph{region \(II\) describes the antipodes of region \(I\)}. The antipodes are found by replacing the solid angle \(\W=(\tht,\,\vv)\) by \(\tl\W=(\pi-\tht,\,\vv+\pi)\).
	
Our third point concerns the ``firewall" problem\,\cite{firewall, Mathur}. It was noted that particles ending up in region \(II\) or \(IV\) of the Penrose diagram (see Fig.~\ref{Penrose}), should be entangled with the Hawking particles coming out. This leads to problems of quantum cloning. If the thermal states that were found by Hawking\,\cite{SWH, SWH2, SWH3} in his analysis of the black hole quantum behaviour would be replaced by a pure state, this modification would be so large that a local observer sees his vacuum state replaced by an infinite shower of out-going particles: a firewall. But this was under the assumption that particles in region \(II\) are invisible to te outside observer. If we accept the antipodal identification explained above, then the Hartle Hawking wave function seen by the distant observer, is no longer a thermally mixed state, but a single pure quantum state. Thus, as long as we constrain ourselves to soft particles only, there is no problem. Now what we propose is that if a soft particle becomes hard, through its time evolution, we must replace it by its gravitational footprint among the surrounding soft particles.  This now precisely restores unitarity of the evolution operator. Since our expansion in spherical harmonics turns all our equations into 1+1 dimensional partial differential equations, what happens now, becomes totally transparent. The spherical mode particles act mutually independently. The Fourier transform is unitary, \emph{but it covers both regions \(I\) and \(II\)} so that unitarity actually requires something like the antipodal identification so as to make all components of the wave functions physically observable.\fn{Critical readers may suspect that the antipodal identification of regions \(I\) and \(II\) should have huge physical consequences that invalidate the idea. We emphasise that this is not so. The position operator \(u^\pm(\tht,\vv)\) of a particle tells us it is in region \(I\) when positive, and in region \(II\) when negative, or, \(u^\pm(\W)=-u^\pm(\tl\W)\). This gives us the constraint that,  in the spherical wave expansion, \(\ell\) can only be odd. We cannot put the particles in a spherically symmetric dust shell (\(\ell=0\)). This strange looking restriction, however, applies to the local observer. The distant observer sees the Hawking particles themselves in spherically symmetric states, see Section \ref{antipodes}.}

\newsecl{The basic, explicit, calculation}{calc}

An essential element of our calculation is the Fourier transformation using tortoise (Krus\-kal-Szekeres) coordinates.  Equations \eqn{ellmdef}---\eqn{ellmshifts} show that the out-going particles will have wave functions that are the Fourier transforms of the in-going wave functions. Consider the coordinates \(u\) and \(p\) for a local observer, in a mode with given \(\ell\) and \(m\). These wave functions are generated by the Dirac kets \(|u\ket\) and \(|p\ket\). We have
	\be[\,u,p\,]=i\ , \quad\hbox{ so that }\quad \bra u|p\ket=\frac 1{\sqrt{2\pi}}\,e^{ipu}\ .\ee
A wave function \(|\psi\ket\) is defined by its inner products \(\bra u|\psi\ket\). Its Fourier transform is
	\be \hat\psi(p)\equiv\bra p|\psi\ket=\frac 1{\sqrt{2\pi}}\int_{-\infty}^\infty\dd u\,e^{-ipu}\psi(u)\ . \ee	
Now an outside observer will use coordinates that for the local observer look like exponentials. Furthermore, the outside observer (at one side of the black hole) only has access to the positive values of \(u\). The negative values describe the antipodal domain. Therefore, we write
	\be u\equiv \s_u\,\ex{\r_u}\ ,\qquad p\equiv \s_p\,\ex{\r_p}\ ;\qquad \s_u=\pm 1\ ,\qquad \s_p=\pm 1\ . \ee
\(\r_u\) and \(\r_p\) run from \(-\infty\) to \(\infty\)\,. Since \(u\) and \(p\) do not commute, \(\s_u\) and \(\r_u\) do not commute with \(\s_p\) and \(\r_p\)\,.	So, one of these pairs suffice to define the wave function, \(\psi_{\s}(\r)\)\,. Transforming from the variable \(u\) to \(\s_u,\,\r_u\), we write
	\be\tl\psi_{\s_u}(\r_u)\equiv \ex{\halfje\r_u}\,\psi(\s_u\,\ex{\r_u})\ ,\qquad\tl{\hat \psi}i_{\s_p}(\r_p)\equiv\,\ex{\halfje\r_p}\,\hat\psi (\s_p\,\ex{\r_p})\ ;\ee
	the exponents in front are there to ensure that the wave functions in the new coordinates are properly normalised. Working out the integrals, one gets
	\be \tl{\hat\psi}_{\s_p}(\r_p)=\sum_{\s_u=\pm 1}\int_{-\infty}^\infty\dd\r_u\,K_{\s_u\s_p}(\r_u+\r_p)\,\tl\psi_{\s_u}(\r_u) \ ,\nm \\
	\hbox{with } \  K_\s(\r)\equiv\frac 1{\sqrt{2\pi}}\,\ex{\halfje\r}\,\ex{-i\s\,e^\r}\ . \eel{newkernel}
	
Now the kernel for the Fourier transform, \(\ex{ipu}\) is entirely non-local, but in terms of the exponentiated coordinates, locality is restored to some extent. The kernel \eqn{newkernel}  drops exponentially for negative \(\r\) and oscillates so fast  for large positive values of \(\r\) that its effects also drop very fast for positive \(\r\), when convoluted with any sufficiently smooth test function.

Notice the symmetry under \(\r_u\ra\r_u+\l\ ,\ \ \r_p\ra\r_p-\l\ ,\) which is the symmetry \(u\ra u\,e^\l\ , \ \ p\ra p\,e^{-\l}\ ,\) a property of the Fourier transform, here just reflecting invariance under time translations.  We now use this symmetry to write the states	as energy eigen states. Let \(\k\) be the energy in the scaled variables (\(\k=4GME\), see Eq.~\eqn{timescale})\,.
	Near the horizon, the Hamiltonian is the dilation operator,
	\be H=-\half(u^+\,p^-+p^-\,u^+)=\half(u^-p^++p^+u^-)\  =\nm\\
	i\frac{\pa}{\pa\r_{u^+}}=-i\frac\pa{\pa\r_{u^-}}=-i\frac\pa{\pa\r_{p^-}}=i\frac\pa{\pa p_{\r^+}}=\k\ ,
	\eel{ham}
 The energy eigen states are then \(C(p^-)^{i\k}\ra C_\s \ex{i\k\,\r_{p^-}}\,.\).  We now need the Fourier transform of the kernel \eqn{newkernel}, using the integration variable \(y=e^{\r}\):
\be F_\s(\k)=\frac 1{\sqrt{2\pi}}\int_0^\infty\frac{\dd y}y y^{\halfje-i\k}\,e^{-i\s y}\iss\frac 1{\sqrt{2\pi}}\G(\half-i\k)\ex{-\fract
{i\s\pi}4-\fract\pi 2\k\s}\ , \eel{EulerGamma}
where, again, \(\s=\pm 1\).

The summation in Eq.~\eqn{newkernel} implies that the kernel is now a \(2\times 2\) matrix, which, by construction, must be unitary. Here, we see this explicitly:
	\be \begin{pmatrix} F_+&F_-\\ F_-&F_+ \end{pmatrix} \ \hbox{ is unitary since } \ F_+F^*_-=-F_-F^*_+ \ 
	\hbox{ and } \ |F_+|^2+|F_-|^2=1\ ,\ee
which follows from a well-known property	of the Euler Gamma function:
	\be \G(x)\G(1-x)=\frac \pi{\sin\pi x}\ . \ee

From Eq.~\eqn{ellmshifts}, we now derive the unitary evolution operator relating the momentum distribution of the in-states to that of the out-states, after inserting the scale factor \(\frac{8\pi G}{\ell^2+\ell+1}\). If \(u^\pm=\s_\pm\ex{\r^\pm}\), then
	\beq \psi^\inn_{\s_+}\,\ex{-i\k\r^+}&\ra \psi^\outt_{\s_-}\,\ex{i\k\r^-}\ ,& \\
	\psi^\outt_{\s_-}&=\sum_{\s_+=\pm 1}F_{\s_+\s_-}(\k)\,
	\ex{-i\k\log\left(8\pi G/(\ell^2+\ell+1)\right)}\,\psi_{\s_+}^\inn\ . \eeql{inoutmatrix}
These equations generate the contribution to the evolution operator from all \((\ell,\,m)\) sectors of the system, 
where \(|m|\le\ell\). At every \((\ell,\,m)\), we have a contribution to the position operators \(u^\pm(\tht,\,\vv)\) and 
momentum operators \(p^\pm(\tht,\,\vv)\) proportional to the spherical harmonics \(Y_{\ell m}(\tht,\,\vv)\). 
The signs \(\s_\pm\) of \(u^\pm(\tht,\,\vv)\) tell us whether we are in region \(I\) or region \(II\). 
The signs of \(p^\pm(\tht,\,\vv)\) tell us whether we added or subtracted a particle from region \(I\) or region \(II\).

\begin{figure} [h]\qqquad\quad
\lowerwidthfig{0pt}{320pt}{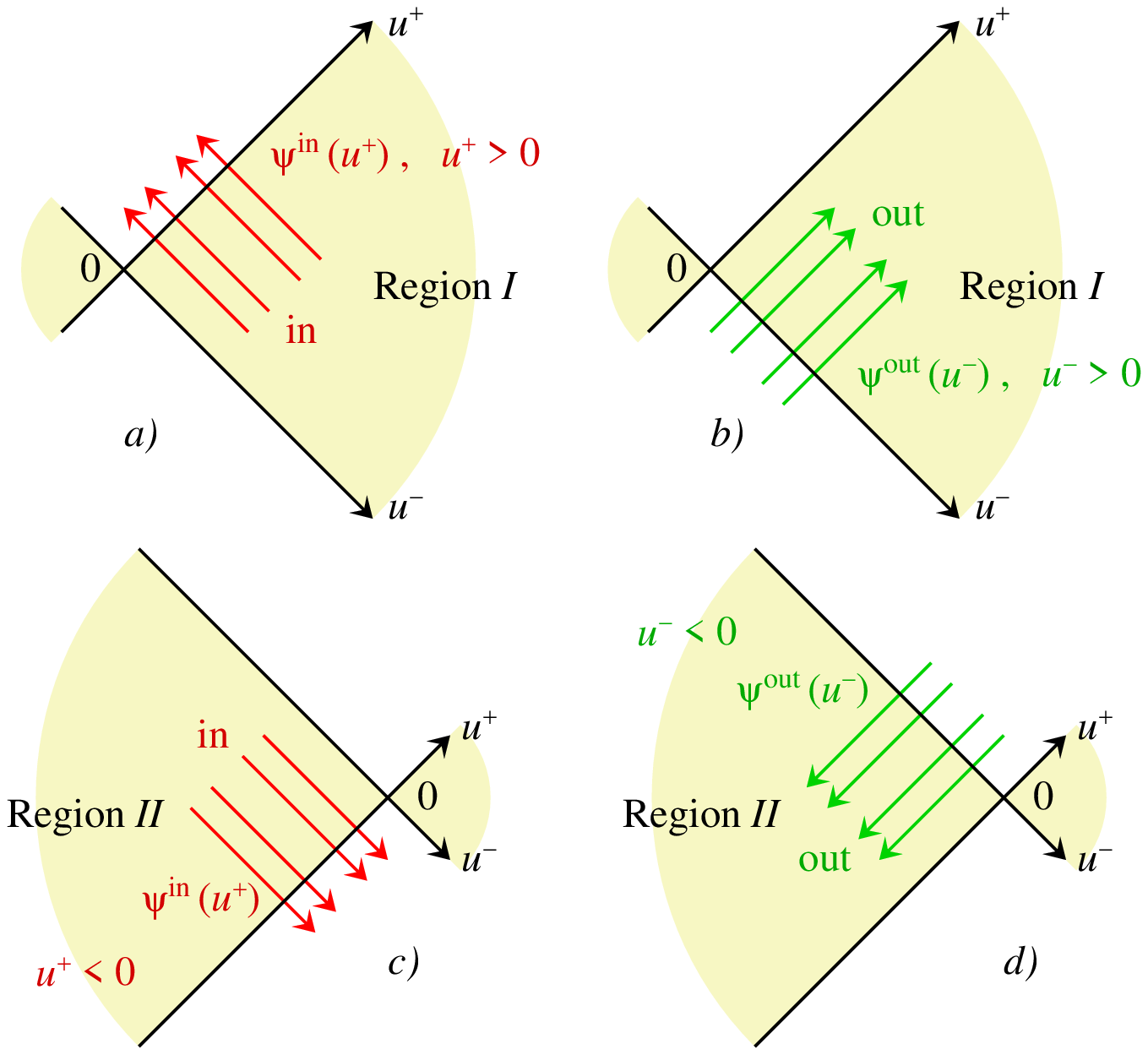} \quad \begin{caption}
{ \small Particles moving in and out, in regions \(I\) and \(II\). $a)$ Wave functions \(\psi(u^+)\) of particles moving in in region \(I\), have \(u^+>0\). $b)$ Out-particles have wave functions \(\psi(u^-)\) with \(u^->0\). $c,\,d)$ In region \(II\), in-particles have \(u^+<0\) and out-particles have \(u^-<0\).
 \labell{regions}}
 \end{caption} \end{figure}

The locations of in-going and out-going particles in regions \(I\) and \(II\) of the Penrose diagram (or the Kruskal-Szekeres coordinates) are depicted in Fig.~\ref{regions}. Note that the in-particles will never get the opportunity to become truly hard particles, because as soon as they become hard, we replace their wave functions by their footprints in the wave functions of the out-particles, \(p^-_\inn\ra u^-_\outt\). When \(p^-\) gets \emph{very} large, the \(u^-\) coordinate of the out-particle becomes large. Thus, the out-particle simply leaves the system. Similarly, the out-particles at very early times \(\t\), would carry large values for \(p^+\); again. we replace them by the wave functions of the in-particles, which are still far away (\(u^+\) is still large, and \(p^-\) is still very small. Particles that are far away are taken to be outside our system altogether.

This is how we get rid of the `firewalls', which for a long time were ill-understood.\cite{firewall}

Regions \(III\) and \(IV\) in the Penrose diagram (Fig.~\ref{Penrose}) never play much of a role. At best, we can sort the particles living there in domains beyond \(t\ra\infty\) or \(\t\ra\-\infty\), so they occupy distinct places of the time coordinate, and consequently there is no need to worry about quantum cloning. For the physics of a black hole at finite time (for the distant observer) we may exclude these particles from consideration.

There are some important warning signs however.  Our procedure has given us the unitary evolution operator, but it is formulated in terms of the \emph{total momentum distribution} \(p^\pm(\tht,\,\vv)\), and the \emph{centre-of-mass positions} \(u^\pm(\tht,\,\vv)\) only. We would need the quantum wave functions as \ul{elements of Fock space} (which would be specified by individual positions or momenta of all particles, as well as other quantum numbers than the geometrical ones). The exact procedure for doing this right has not yet been worked out, but we find the situation to be quite analogous to what we have in string theory amplitudes\,\cite{GtHstrings1, GtHstrings2}; in string theory, the in- and out-going particles are represented as vertex insertions. Here also, only the total momentum counts, but the integration over the vertex positions on the world sheet at the end do give is the individual n-particle states. These amplitudes, at the time, were needed to explain the experimental observations, so we clearly observe how important the experiments have been in this field, and how easy it is to make mistakes when experimental clues are absent.

An other, probably related warning sign is, that we have too many ``microstates". It seems that the total number of Hawking particles involved in a black hole creation and evaporation process is finite, roughly of order \(M_\BH^2\) in Planck units. This will mean in practice that a limit \(\ell<\ell_{\mathrm{max}}\) must be imposed, with \(\ell_{\mathrm{max}}=\OO(M_\BH)\). This is precisely where the gravitational  forces in the transverse direction come in, so that all particles in this domain will tend to become hard. How exactly to impose a constraint here is also not yet well understood.

Without claiming that we resolved all difficulties, we do emphasise that many of the `problems' and `paradoxes encountered in Refs\,\cite{firewall, Giddings1, Polch, Giddings4} and others, do not show up or are resolved  in the treatment given here. The scheme described here was first clearly exposed in Ref.\,\cite{GtHrecent1},

\newsecl{The antipodal identification}{antipodes}

Regions \(I\) and \(II\) are exact copies of one another, while the soft particles they contain will in general differ. Often it was thought that region \(II\) describes something like the `inside' of the black hole. More daringly, it is also often taken to be a different black hole, somewhere else in our universe,\fn{\emph{where} in our universe? Since the outside world does not show any sort of connection to another black hole, this question would be totally unanswerable, and in fact would form an argument against this view.}  or in some other universe. These various options do not differ very much; in all these cases, the interior would be some arbitrary, unaccessible domain of science.

The important thing here is that region \(II\) also has an asymptotic region. The evolution matrix derived in the previous section involves both asymptotic domains. The Fourier transform of a wave function of an in-going particle in region \(I\) must cover both domains, as demanded by analyticity. This means that the evolution operator strongly involves features in both domains; we cannot ignore domain \(II\). If experimental data were available, the question as to what region \(II\) represents would be an urgent topic of research. Now we have no access to such data, and all we can do is argue as accurately as possible what our options are.

A black hole must be assumed to act just like any other form of matter: whenever it is brought in contact with whatever outside forms of matter, it should interact with it. All forms of matter that we know react by way of a unitary scattering matrix. In objects as large as black holes, it will be more appropriate to talk of an evolution operator, since the truly asymptotic states may be separated by many times the lifetime of the universe, and it would not be practical to force us to wait that long. We need to know the black hole responses in shorter lapses of time. The evolution operator derived in the previous sector does just this, but it does not disclose what region \(II\) really is. Now the options turned out actually to be quite limited.

The above already left us with one general conclusion: region \(II\) should also represent the same black hole as region \(I\) does. However, if region \(II\) would describe the same spots on the horizon as region \(I\), this would generate a cusp singularity on the horizon, and interfere with our ability to use the data of a local observer to work out what happens. What is needed is a mapping \(A\) bringing us to another point on the horizon: \(\tl\W=A(\W)\). What is \(A\)?
\(A\) must be an isometry of the horizon, so \(A\in O(3)\). Then, since there are no more than two regions, \(A^2=\mathbb I\). Therefore, its eigen values are \(\pm 1\). If any of the three eigen values would be \(+1\), there would be a fixed point, where we would encounter our troublesome cusp singularity. 

Therefore, all eigen values are \(-1\), which implies that \(A=-\mathbb I\), the antipodal mapping. To be precise, we have here the mapping \((\tht,\,\vv)\leftrightarrow (\pi-\tht,\,\vv+\pi)\). Antipodal \emph{identification} only holds for the central point (the origin) of the Penrose diagram. That region, normally having the topology of an \(S_2\) sphere, now is seen to be a \emph{projective sphere}. But relating region \(I\) with the antipodes of region \(II\) means that the mapping from Schwarzschild coordinates to Kruskal-Szekeres coordinates is \ul{one-to-one}, rather than one-to-two (that is, \((r,\,t)\ra (x,\,y)\) \emph{and} \((-x,\,-y)\). This means that we arrived at a new constraint imposed on all general coordinate transformations: 

\begin{quote} \emph{ In  applying general coordinate transformations for quantised fields on a curved space-time background, to use them as a valid model for a physical quantum system, one must demand that the following constraint hold: the mapping must be one-to-one and differentiable.} \end{quote}
In particular, this must apply to the asymptotic regions of the surrounding (almost) flat space-time.
The emergence of a non-trivial topology needs not be completely absurd, as long as no signals can be sent around in loops. This is the case at hand here. Requiring the absence of singularities in the physical domain of space- time forces us to the antipodal folding.

To illustrate that the antipodal mapping is topologically non-trivial, we point out, in the presence of the black hole space-time,  the existence of M\"obius strips such that, in closing the path along the strip, the arrow of time is inverted. This shows up if we follow a (space-like) path through the horizon, closing it by travelling \(180^\circ\) to the antipode outside the black hole. The arrow of time then inverts while crossing the horizon.

Now only a time reversal \(T\) takes place along this M\"obius strip, while \(P\) and \(C\) may be conserved. So we have a \(CPT\) inversion along the loop. Since the laws of nature are invariant under \(CPT\) there is no clash with quantum field theory here.\fn{This point raised some confusion in earlier discussions, but it is now cleared.}

In discussions, a question arose concerning the time spent by a particle absorbed by a black hole, before it is re-emitted as Hawking radiation. Since our scattering matrix is very similar to the one that would be obtained if we started from a \emph{brick wall} at a Planckian distance from the horizon, we expect the answer to be similar: \emph{For the external observer, a particle takes some time to reach the brick wall, it then bounces, and takes about the same amount of time to leave the black hole environment.} This amount of time is \(\OO(M_\BH\log M_\BH)\) in Planck units. For the black hole scattering operator \eqn{inoutmatrix}, the same result applies; although the integration kernel \(K_\s\) in \eqn{newkernel} does not have a compact support, its range is limited just like that of the brick wall if we let it act on any sufficiently smooth test function.

But, in as far as the question is well-put, other answers may be considered. The brick wall generates a deep potential well, which may actually hold particles for a much longer amount of time. If we assume thermal equilibrium to arise, at the Hawking temperature, we see that many of the Hawking particles will stay trapped since they carry not enough kinetic energy to escape. We see an \emph{atmosphere} of particles near the horizon. These particles might hang around for time scales much longer than \(M_\BH\) in Planck units. Who is right?

Our scheme gives the following mathematical answer: of all quantum states of the black hole, there is only one stationary state, the Hartle Hawking state. This state describes a completely entangled situation of particles streaming into the black hole and particles leaving, forming a perfect equilibrium. The black hole is in a heat bath of Hawking particles. These particles are completely entangled, so that we cannot identify individual particles at all. For any individual particle, the time spent near the horizon is an ill-defined concept; it could be set to infinity.

The Hartle-Hawking state, here a pure quantum state, coincides with the vacuum state for a local (freely falling) observer. All other states a black hole can be in, are obtained from the HH state by applying creation operators to it, as seen by the local observer. The global observer will see these operators as superpositions of creation and annihilation operators. Thus, at any moment in time, we can modify the state, to see what happens. The local observer will see the extra particles come and go very quickly. In short: any \emph{fluctuations} away from the steady Hartle-Hawking stream of particles will not last for much longer than \(M_\BH\log M_\BH\) in Planck units. This is the best answer we can give to the question.

\vskip15pt 
 \noindent{\large \textbf{Acknowledgement}}\\

The author acknowledges discussions with P.~Betzios, R.~Brustein, G.~Dvali,  A.~Franzen,  N.~Gaddam,  S.~Giddings, S.~Hawking, J.~Maldacena, S.~Mathur, S.~Mukhanov, O.~Papadoulaki, F.~Scardigli, L.~Susskind, and W.~Vleeshouwers among others.

\end{document}